\documentclass[sigconf]{acmart}

\usepackage{multirow}
\usepackage{algorithm}
\usepackage{algpseudocode}

\AtBeginDocument{
  }

\setcopyright{acmlicensed}
\copyrightyear{2018}
\acmYear{2018}
\acmDOI{XXXXXXX.XXXXXXX}
\acmConference[Conference acronym 'XX]{Make sure to enter the correct
  conference title from your rights confirmation email}{June 03--05,
  2018}{Woodstock, NY}
\acmISBN{978-1-4503-XXXX-X/2018/06}

\begin{document}

\title{Distribution-Aware End-to-End Embedding for Streaming Numerical Features in Click-Through Rate Prediction}

\author{Jiahao Liu}
\authornote{Work completed during Jiahao Liu and Hongji Ruan's internships at Tencent. Hongji Ruan co-designed the system and led the online experiments.}
\affiliation{
  \institution{Fudan University}
  \city{Shanghai}
  \country{China}
}
\email{jiahaoliu23@m.fudan.edu.cn}

\author{Hongji Ruan}
\authornotemark[1]
\affiliation{
  \institution{Beijing Jiaotong University}
  \city{Beijing}
  \country{China}
}
\email{23125251@bjtu.edu.cn}

\author{Weimin Zhang}
\authornote{Corresponding author.}
\affiliation{
  \institution{Tencent}
  \city{Beijing}
  \country{China}
}
\email{weiminzhang@tencent.com}

\author{Ziye Tong}
\affiliation{
  \institution{Tencent}
  \city{Beijing}
  \country{China}
}
\email{ziyetong@tencent.com}

\author{Derick Tang}
\authornotemark[2]
\affiliation{
  \institution{Tencent}
  \city{Beijing}
  \country{China}
}
\email{dericck@tencent.com}

\author{Zhanpeng Zeng}
\affiliation{
  \institution{Tencent}
  \city{Guangzhou}
  \country{China}
}
\email{marcuszeng@tencent.com}

\author{Qinsong Zeng}
\affiliation{
  \institution{Tencent}
  \city{Guangzhou}
  \country{China}
}
\email{qinzzeng@tencent.com}

\author{Peng Zhang}
\affiliation{
  \institution{Fudan University}
  \city{Shanghai}
  \country{China}
}
\email{zhangpeng\_@fudan.edu.cn}

\author{Tun Lu}
\authornotemark[2]
\affiliation{
  \institution{Fudan University}
  \city{Shanghai}
  \country{China}
}
\email{lutun@fudan.edu.cn}

\author{Ning Gu}
\affiliation{
  \institution{Fudan University}
  \city{Shanghai}
  \country{China}
}
\email{ninggu@fudan.edu.cn}

\renewcommand{\shortauthors}{Jiahao Liu et al.}

\begin{abstract}
This paper explores effective numerical feature embedding for Click-Through Rate prediction in streaming environments. Conventional static binning methods rely on offline statistics of numerical distributions; however, this inherently two-stage process often triggers semantic drift during bin boundary updates. While neural embedding methods enable end-to-end learning, they often discard explicit distributional information. Integrating such information end-to-end is challenging because streaming features often violate the i.i.d. assumption, precluding unbiased estimation of the population distribution via the expectation of order statistics. Furthermore, the critical context dependency of numerical distributions is often neglected. To this end, we propose DAES, an end-to-end framework designed to tackle numerical feature embedding in streaming training scenarios by integrating distributional information with an adaptive modulation mechanism. Specifically, we introduce an efficient reservoir-sampling-based distribution estimation method and two field-aware distribution modulation strategies to capture streaming distributions and field-dependent semantics. DAES significantly outperforms existing approaches as demonstrated by extensive offline and online experiments and has been fully deployed on a leading short-video platform with hundreds of millions of daily active users.
\end{abstract}



\maketitle

\begin{figure}[t]
  \centering
  \includegraphics[width=0.95\linewidth]{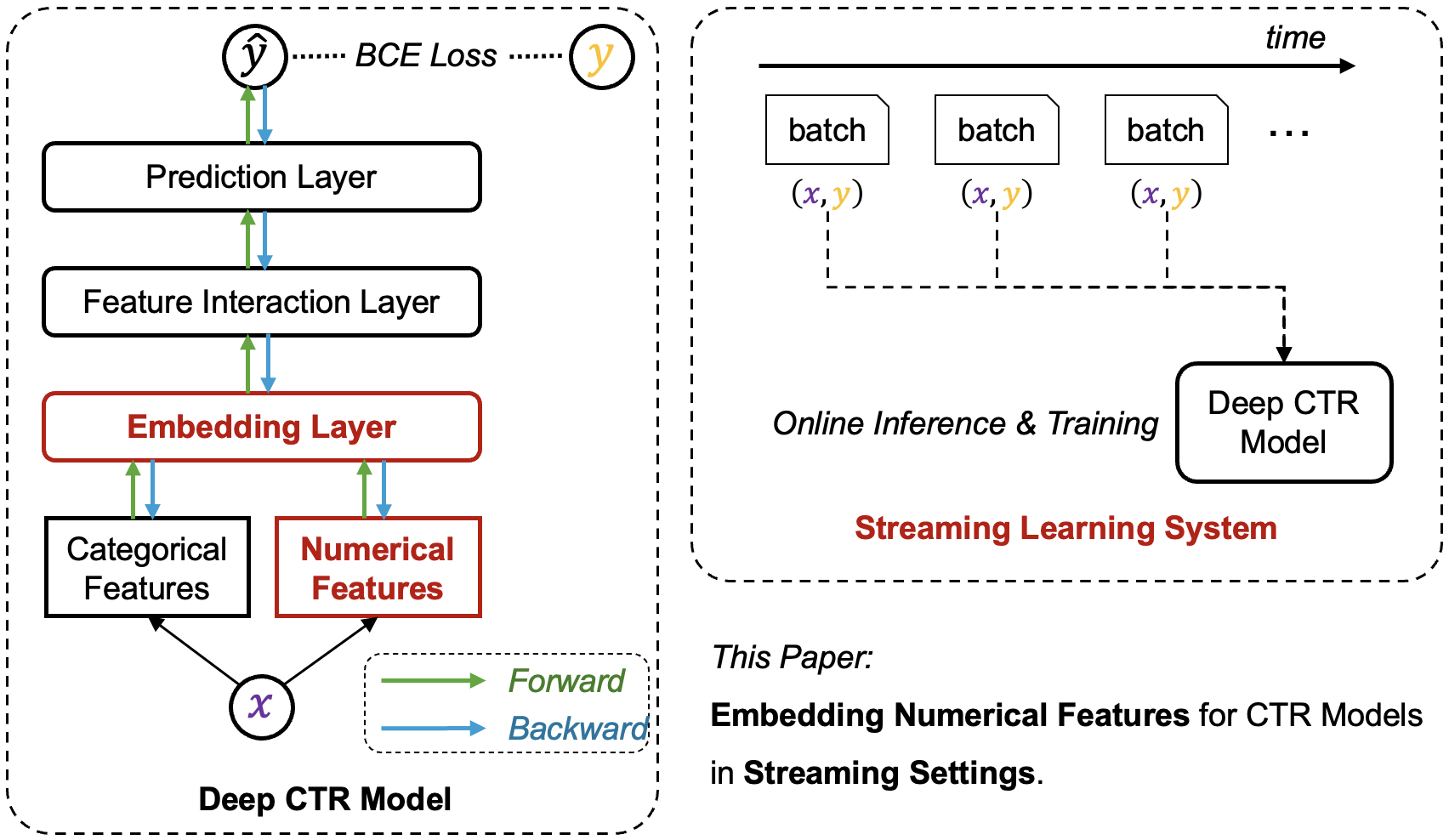}
  \caption{Illustration of the Deep CTR model architecture and the streaming learning system. This paper explores effective numerical feature embedding for CTR prediction in streaming environments.}\label{fig:9mxk}
  \Description{}
\end{figure}

\section{Introduction}

\begin{figure*}[t]
  \centering
  \includegraphics[width=0.95\linewidth]{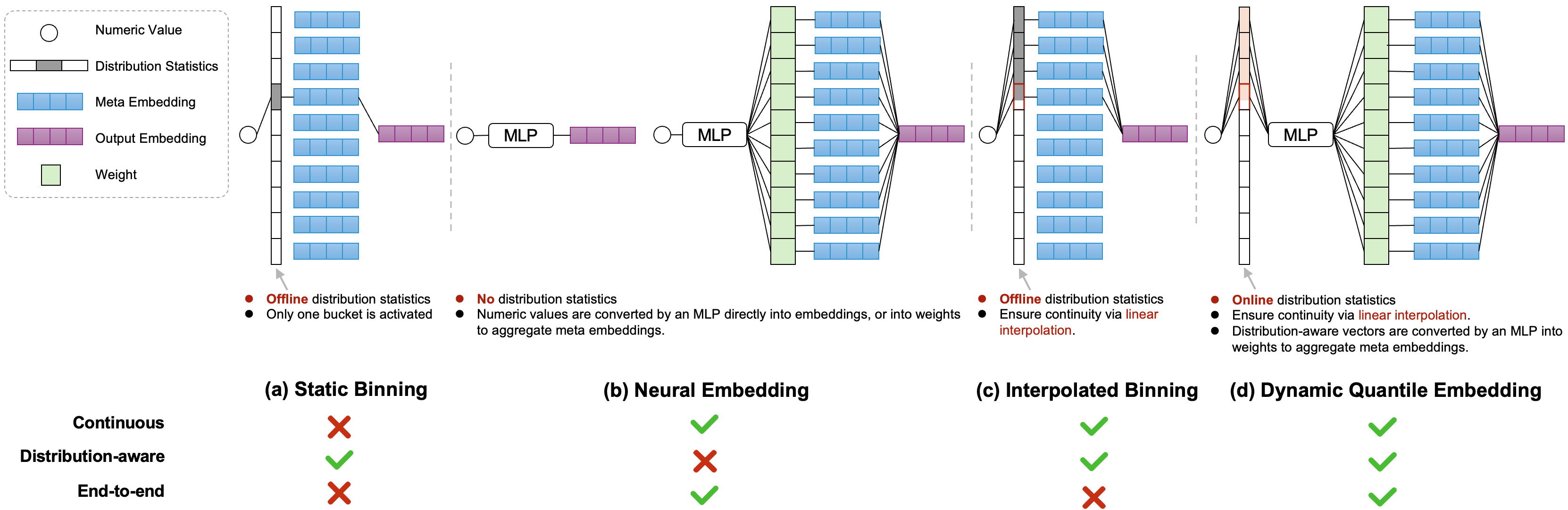}
  \caption{Comparison of different numerical feature embedding paradigms. (a) Static Binning: mapping values to discrete buckets via offline statistics. (b) Neural Embedding: transforming values through differentiable neural layers. (c) Interpolated Binning: aggregating meta-embeddings via linear interpolation with static buckets. (d) Dynamic Quantile Embedding: generating distribution-aware weights using online-estimated quantiles.}\label{fig:admo}
  \Description{}
\end{figure*}
Click-Through Rate (CTR) prediction aims to estimate the probability of a user clicking a specific item, functioning as a pivotal engine in content recommendation and computational advertising systems~\cite{richardson2007predicting,liu2022parameter,liu2023personalized}. Input features are typically categorized into categorical features (e.g., user gender and item category) and numerical features (e.g., user age and item price). Most mainstream deep CTR models follow a canonical architecture comprising an embedding layer, a feature interaction layer, and a prediction layer. Specifically, all features are first projected into a low-dimensional embedding space, after which high-order relationships are captured via feature interaction layers to produce the final prediction. As the model's cornerstone, the embedding layer is parameter-intensive, accounting for the bulk of the total parameters and dictating overall predictive performance~\cite{joglekar2020neural,zhang2016deep,liu2023triple}. For categorical features, a lookup-table mechanism is standard practice for mapping each discrete value to a unique embedding vector~\cite{guo2017deepfm,qu2018product,liu2023recommendation}.

Concurrently, numerical features represent a vital information source extensively leveraged in industrial applications~\cite{guo2021embedding,liu2023autoseqrec,han2025fedcia}. In particular, statistical features derived from profound business insights often yield substantial online performance gains~\cite{guo2021online,gu2025llm,liu2025filtering}. However, unlike their categorical counterparts, numerical features pose challenges for simple lookup-table operations due to the presence of continuous distributions and inherent ordinality~\cite{shen2023dae,han2025fedcia}. Furthermore, to accommodate real-time shifts in user interests and item popularity, models are typically trained within a \textbf{streaming learning paradigm}. In this setting, data arrive continuously in a batch-wise manner, necessitating online model updates to capture evolving distributions\footnote{In industrial practice, CTR models commonly exhibit a pronounced “one-epoch” phenomenon, where the model converges to optimal performance within the first training epoch, and additional epochs lead to overfitting. Consequently, models are typically trained in a single-pass fashion, where each instance is utilized for parameter updates exactly once~\cite{zhang2022towards}.}. As illustrated in Figure \ref{fig:9mxk}, \textbf{this paper explores effective numerical feature embedding for CTR prediction in streaming environments}.

A straightforward approach to numerical feature embedding is \textbf{static binning}, which partitions continuous values into discrete bins based on the offline-computed empirical distribution of features and assigns a learnable embedding vector to each bin\footnote{Embedding numerical features resolves the representation space inconsistency between numerical and categorical features, enabling unified feature interaction modeling~\cite{cheng2016wide,liu2025agentcf++,gu2020deep}. We thus focus on embedding-based methods and discuss non-embedding alternatives (e.g., raw values or transformations) in our experiments.} \footnote{Equal-frequency discretization (EFD) is more adaptive to feature distributions and robust to outliers. In contrast, logarithmic-based discretization~\cite{juan2016field} is typically effective only for right-skewed distributions, while equal-width discretization often requires domain expertise~\cite{shen2023dae}. Accordingly, this paper focuses on EFD and provides comparative evaluations with other discretization methods in our experiments.}~\cite{qu2018product}, as illustrated in Figure \ref{fig:admo}(a). This method, however, suffers from two major limitations. First, the encoding is intrinsically discontinuous: numerically distant values within the same bin share identical embeddings, whereas adjacent values across bin boundaries are represented differently, effectively resulting in a step-function mapping. Second, as data distributions evolve over time, bin boundaries must be updated offline. These updates could potentially shift the bin indices of samples, thereby inducing semantic drift.

\textbf{Neural embedding} methods map numerical features to embeddings through differentiable neural architectures, either by directly transforming values or by generating weights to aggregate meta-embeddings~\cite{guo2021embedding,liu2025improving,liu2025unbiased}, as shown in Figure \ref{fig:admo}(b). These methods are inherently continuous and end-to-end trainable, ensuring that embeddings vary smoothly with input values while bypassing the drift inherent in the two-stage pipeline of offline statistics and model training. However, they often overlook the underlying distributional characteristics of numerical features. In contrast, binning-based methods explicitly incorporate distributional priors into the encoding process. Such distribution awareness is crucial for effective and stable representation learning and remains robust across diverse feature distributions.

Explicitly incorporating numerical feature distributions into end-to-end frameworks remains nontrivial, primarily because distribution modeling typically hinges on precomputed offline statistics. For instance, \textbf{interpolated binning} methods require offline equal-frequency quantiles and aggregate meta-embeddings based on these static buckets~\cite{gorishniy2022embeddings,zhang2025evalagent,wu2025bidirectional}, as illustrated in Figure \ref{fig:admo}(c). Although linear interpolation maintains representation continuity, the reliance on offline statistics compromises end-to-end integrity and inevitably introduces semantic drift as bucket boundaries evolve.

To the best of our knowledge, DAE~\cite{shen2023dae} is the first and only approach that models distributional information in an end-to-end manner under stringent memory and computational constraints. We refer to this class of approaches as \textbf{dynamic quantile embedding}. Specifically, DAE leverages online-estimated quantiles to encode numerical features with distributional knowledge; it then employs a neural network to transform these distribution-aware vectors into weights for aggregating meta-embeddings, as illustrated in Figure \ref{fig:admo}(d). DAE's quantile estimation hinges on the strong assumption that data from successive training batches are independently and identically distributed (i.i.d.). Under this condition, it is provable that the expectation of order statistics yields an unbiased estimate of the quantiles. However, this assumption is frequently violated in streaming training scenarios\footnote{In Appendix \ref{appendix:993k}, we analyze the non-stationarity of the two public datasets used in our experiments.}. In the presence of non-stationary feature distributions, estimation can suffer from systematic bias, thereby undermining the stability and efficacy of representation learning\footnote{In Appendix \ref{appendix:93km}, we provide theoretical limitations of order statistics in non-IID streams.}. Consequently, \textbf{effectively incorporating distributional information in an end-to-end manner under streaming conditions remains a critical bottleneck for numerical feature embedding}.

On the other hand, numerical features exhibit significant context-dependency in their semantics: identical values may represent distinct distributions across different subpopulations. For instance, a fixed price point signifies different quality tiers across product categories. However, existing numerical embedding methods typically ignore such conditional distributional shifts, mapping values into context-invariant representations and leaving subsequent feature interaction layers to implicitly handle the induced biases. This approach propagates biased representations throughout the model, undermining the embedding layer's representational capacity and increasing the optimization burden on downstream layers. A straightforward solution is to model numerical distributions independently for each field, integrating field-specific distributional priors during encoding. However, the combinatorial nature of features causes an exponential explosion in marginal distributions when multiple fields intersect, compromising both scalability and flexibility. Consequently, \textbf{adaptively integrating field-aware distributional information emerges as a critical challenge for numerical feature embedding}.

\begin{table}[t]\small
\centering
\caption{Comparison between DAE and DAES.}\label{tab:9cms}
\begin{tabular}{cccc}
\toprule
Method & Streaming & Encoding Space & Field-aware \\
\midrule
DAE & $\times$ & Value & $\times$ \\
DAES (Ours) & $\checkmark$ & Quantile & $\checkmark$ \\
\bottomrule
\end{tabular}
\end{table}

In this work, we introduce \textbf{DAES}, a numerical feature embedding framework designed for streaming scenarios. DAES enables smooth embedding representations and end-to-end training while explicitly incorporating distributional information and performing adaptive distribution modulation. Specifically, to estimate distributions in streaming settings, we develop an efficient method based on reservoir sampling enhanced with a jump sampling mechanism. Unlike DAE~\cite{shen2023dae} which encodes values directly in the value space, we perform distribution-aware encoding in the quantile space. This design explicitly integrates distributional characteristics and mitigates adverse effects of numerical density\footnote{In Appendix \ref{appendix:00kd}, we discuss limitations of value-space encoding.}. To capture the context-dependent nature of numerical features while avoiding combinatorial explosion, we propose two field-aware adaptive distribution modulation mechanisms to explicitly address semantic bias during the embedding stage: (i) a gating network for adaptive encoding modulation, and (ii) a parameter generator producing field-specific weights for conditional refinement. Table \ref{tab:9cms} illustrates the improvements achieved by DAES over DAE. Comprehensive offline experiments validate our design, while large-scale online experiments show DAES significantly outperforms state-of-the-art methods in real-world advertising scenarios. Beyond accuracy gains, DAES eliminates separate preprocessing by seamlessly integrating feature processing into model training, reducing engineering complexity. As a pluggable module compatible with various deep CTR architectures, \textbf{DAES is deployed on a platform serving hundreds of millions of daily active users}.

The main contributions of this work are summarized as follows:
\begin{itemize}
\item To the best of our knowledge, this is the first work to introduce distributional information for numerical feature embedding in streaming scenarios in an end-to-end manner, and to systematically implement a distribution modulation mechanism for numerical distributions.
\item We propose DAES, an efficient distribution estimation method based on reservoir sampling, coupled with field-aware distribution modulation strategies to capture both global streaming distributions and field-dependent semantics.
\item Extensive offline and online experiments demonstrate that DAES significantly outperforms existing approaches and has been adopted in a real-world platform with hundreds of millions of daily active users.
\end{itemize}

\section{Related Work}\label{sec:mmeq}

\subsection{Click-Through Rate Prediction}
To improve the predictive performance of CTR models, how to effectively characterize informative feature interactions has long been a central research topic. Early approaches mainly relied on shallow models, such as logistic regression (LR)~\cite{richardson2007predicting} and various factorization machine (FM)-based models~\cite{blondel2016higher,juan2016field,rendle2012factorization}. These methods are generally limited to modeling low-order feature combinations. With the development of deep learning, researchers began to leverage deep networks to automatically mine high-order feature interaction patterns, thereby significantly enhancing model expressiveness and predictive accuracy. According to how explicit and implicit interaction modules are combined within the network architecture, these methods can be broadly categorized into two paradigms: hierarchical (stacked) architectures and parallel architectures~\cite{chen2021enhancing,wang2022enhancing}.

\textbf{Hierarchical architectures.}
Hierarchical models typically introduce dedicated interaction modules on top of embedding representations to explicitly model feature combinations, followed by deep neural networks (e.g., DNNs) to further learn higher-level implicit interaction information. For explicit interaction modeling, prior work has proposed a variety of structures, including inner-product- or outer-product-based interactions (e.g., PNN~\cite{qu2018product}, ONN~\cite{yang2020operation}), Hadamard-product-based interaction mechanisms (e.g., FM~\cite{rendle2012factorization}, FFM~\cite{pan2018field}), and different forms of cross networks (e.g., CN~\cite{wang2017deep}, CNV2~\cite{wang2021dcn}, XCrossNet~\cite{yu2021xcrossnet}). In addition, NFM~\cite{he2017neural} models interactions through a bi-interaction structure, while attention mechanisms have been introduced to enhance interaction selectivity in models such as AFM~\cite{xiao2017attentional}, AutoInt~\cite{song2019autoint}, DCAP~\cite{chen2021dcap}, and DIEN~\cite{zhou2019deep}. After explicit interaction features are generated, the DNN module applies nonlinear transformations to these intermediate representations to capture more complex and deeper implicit feature relationships.

\textbf{Parallel architectures.}
Parallel models adopt multi-branch network structures to model feature interactions from different perspectives simultaneously. They typically place explicit and implicit interaction networks side by side and fuse them at higher layers. Representative examples include Wide \& Deep~\cite{cheng2016wide}, DeepFM~\cite{guo2017deepfm}, DCN~\cite{wang2017deep}, DCN-V2~\cite{wang2021dcn}, xDeepFM~\cite{lian2018xdeepfm}, and AFN+~\cite{cheng2020adaptive}. In these models, the DNN remains the core component for modeling implicit high-order interactions, while the main differences lie in how explicit interactions are modeled. For instance, Wide \& Deep uses LR as the “wide” component to strengthen memorization of historical co-occurrence patterns; DeepFM leverages the FM~\cite{rendle2012factorization} structure to adaptively learn second-order feature interactions; DCN and DCN-V2 propose different forms of cross networks (CN and CN-V2) to efficiently extract feature interactions with bounded order; and xDeepFM introduces the CIN module to capture more complex bounded-order combinational relationships. Beyond these, some studies further enhance model expressiveness by jointly training three or more parallel subnetworks, such as FED~\cite{zhao2020dimension}, NON~\cite{luo2020network}, and MaskNet~\cite{wang2021masknet}.

\subsection{Embedding for Numerical Features in CTR Prediction}
Numerical feature embedding methods can be categorized into four types: static binning, neural embedding, interpolated binning, and dynamic quantile embedding. Unlike static binning, the latter three enable smooth embedding representations with continuous input values.

\textbf{Static Binning (Figure \ref{fig:admo}(a)).} This approach maps features to deterministic buckets based on predefined rules. Typical methods include equal-width (EWD), equal-frequency (EFD)~\cite{qu2018product}, logarithmic-based (LD)~\cite{juan2016field}, and tree-based discretization (TD)~\cite{grabczewski2005feature,ke2019deepgbm,he2014practical}. These methods cannot be optimized end-to-end with respect to the final prediction objective, and typically lack continuity and discriminative power. While MultiHot~\cite{zhangmultihot} mitigates discontinuities by activating neighboring buckets, staircase effects persist.

\textbf{Neural Embedding (Figure \ref{fig:admo}(b)).} Neural embedding facilitates continuous mapping and end-to-end optimization. For instance, YouTube~\cite{covington2016deep} utilizes power transformations, while DLRM~\cite{naumov2019deep} passes features through neural networks. Field Embedding (FE)~\cite{gorishniy2021revisiting}, including DeepFM~\cite{guo2017deepfm} and AutoInt~\cite{song2019autoint}, applies a field-level vector scaled by the numerical value; however, its limited capacity often struggles to capture complex distributions. To enhance representation, AutoDis~\cite{guo2021embedding} employs a differentiable soft discretization module to aggregate meta-embeddings. NaryDis~\cite{chen2022numerical,pan2024ads} integrates mixed-granularity discretizations via attention to improve continuity and discriminability, but it does not explicitly model feature distributions. Other variants utilize B-splines~\cite{shtofffunction} or distance functions~\cite{guo2019multi} for smoothness, while DEER~\cite{cheng2022dynamic} learns quantile boundaries through parameterized bucket widths but lacks distributional supervision.

\textbf{Interpolated Binning (Figure \ref{fig:admo}(c)).} This category achieves smoothness via interpolation while incorporating distribution information. For example, PLE~\cite{gorishniy2022embeddings} leverages precomputed quantiles through cumulative distribution encoding, linear interpolation~\cite{cheng2022dynamic}, or Gaussian interpolation~\cite{zhangmultihot}. Despite using distributional knowledge, their reliance on offline statistics prevents full end-to-end training and may cause semantic drift.

\textbf{Dynamic Quantile Embedding (Figure \ref{fig:admo}(d)).} DAE~\cite{shen2023dae} is currently the only end-to-end approach integrating distributional information. It utilizes order statistics from i.i.d. batch samples as unbiased quantile estimates to guide meta-embedding aggregation. However, in non-stationary streaming scenarios, numerical distances alone may fail to capture evolving distributions, limiting its online efficacy.

\textbf{Other Related Work.} Recent studies also explore feature order and correlation (PCR~\cite{ma2025embedding}, CCSS~\cite{xu2024enhancing}) or automated bucket sizing (AutoLogDis~\cite{yecsil2023numerical}), further highlighting the importance of high-quality numerical embeddings.

\section{DAES}
This section introduces DAES for numerical feature representation. Given a numerical feature, DAES operates through three sequential stages: (1) \textbf{Distribution Estimation} (§\ref{sec:93jf}) employs jump reservoir sampling to efficiently estimate global quantiles from streaming data; (2) \textbf{Quantile Space Interpolation} (§\ref{sec:mmmd}) transforms the input value into a quantile-based coordinate that captures cumulative density information; (3) \textbf{Field-Aware Meta-Embedding Aggregation} (§\ref{sec:mmqu}) modulates the quantile coordinate into a context-aware weight vector based on field embeddings, and aggregates learnable meta-embeddings to produce the final representation.

\subsection{Problem Formulation}
In CTR prediction, each sample is defined as $(\mathbf{x}, y)$, where $\mathbf{x} = [x^{(\mathrm{cat})}_1, \ldots, x^{(\mathrm{cat})}_p; x^{(\mathrm{num})}_1, \ldots, x^{(\mathrm{num})}_q]$ encompasses $p$ categorical and $q$ numerical features. For categorical inputs, we apply a standard embedding look-up $\mathbf{e}^{(\mathrm{cat})}_i = \mathrm{LookUp}(x^{(\mathrm{cat})}_i) \in \mathbb{R}^d$. This paper focuses on the effective representation of numerical features by utilizing a mapping function $F: \mathbb{R} \rightarrow \mathbb{R}^d$ to project scalars into embeddings:
\begin{equation}
\mathbf{e}^{(\mathrm{num})}_i = F(x^{(\mathrm{num})}_i)\text{.}
\end{equation} The final prediction $\hat{y}$ is obtained by passing the concatenated embeddings through feature interaction and prediction layers. The model is optimized by minimizing the binary cross-entropy loss:
\begin{equation}
L = -\frac{1}{N}\sum_{i=1}^{N} \left[ y_i \log(\hat{y}_i) + (1 - y_i)\log(1 - \hat{y}_i) \right] + \lambda \lVert \boldsymbol{\Theta} \rVert_2^2\text{,}
\end{equation}
where $\lambda$ is the regularization coefficient and $\boldsymbol{\Theta}$ represents the parameter set (including the trainable components of $F$).

\subsection{Distribution Estimation over Data Streams}\label{sec:93jf}

\subsubsection{Reservoir Sampling}
Consider a data stream $\mathcal{S} = \{ x_t \}_{t=1}^{\infty}$, where each $x_t \in \mathbb{R}$ is a numerical feature arriving at time $t$. We aim to estimate the $\alpha$-quantile $q_\alpha$ ($\alpha \in (0,1)$) from $\mathcal{S}_t = \{x_1, \dots, x_t\}$ under strict memory constraints that preclude full data retention. Given that non-stationarity and temporal correlations often bias local-statistic approaches (Appendix \ref{appendix:93km}), we employ \textit{reservoir sampling} to maintain an unbiased approximation of the global distribution, from which we derive quantile estimates.

Specifically, we maintain a fixed-size subset $\mathcal{R}_t \subseteq \mathcal{S}_t$ ($|\mathcal{R}_t|=m, m \ll t$). The reservoir is initialized with the first $m$ elements, and each subsequent $t$-th element ($t > m$) replaces a randomly selected member with probability $m/t$. This mechanism guarantees that for any $t \ge m$, each historical sample $x_i \in \mathcal{S}_t$ is retained with a uniform probability:
\begin{equation}\label{eq:d9mg}
P(x_i \in \mathcal{R}_t) = \frac{m}{t}\text{.}
\end{equation}
Implementation details of reservoir sampling and the proof of Equation (\ref{eq:d9mg}) are provided in Appendix \ref{appendix:d9mp} and \ref{appendix:kkvz}, respectively.

\subsubsection{Quantile Estimation}
We formally define the reservoir-based quantile estimator as follows:
\begin{definition}[Reservoir-based Quantile Estimator]
Given a reservoir $\mathcal{R}_t = \{r_1, \dots, r_m\}$ sampled at time $t$, the empirical distribution is $\hat{F}_t(x) = \frac{1}{m} \sum_{j=1}^m \mathbf{1}(r_j \le x)$. The corresponding $\alpha$-quantile estimator is defined as $\hat{q}_\alpha^{(t)} = \inf \{ x \in \mathbb{R} : \hat{F}_t(x) \ge \alpha \}$.
\end{definition}
Eq. (\ref{eq:d9mg}) ensures that $\mathcal{R}_t$ is a uniform sample without replacement from $\mathcal{S}_t$, rendering $\hat{F}_t(x)$ an unbiased estimator of $F_t(x)$. Per the Glivenko–Cantelli theorem, $\hat{F}_t$ converges uniformly to $F_t$. Since the quantile function is the generalized inverse of the CDF, this uniform convergence implies estimator consistency:
\begin{theorem}[Consistency]\label{th:93mv}
Let $q_\alpha^{(t)}$ be the true $\alpha$-quantile of $F_t(x)$. As $m \to \infty$, $\hat{q}_\alpha^{(t)}$ converges in probability to $q_\alpha^{(t)}$: $\hat{q}_\alpha^{(t)} \xrightarrow{p} q_\alpha^{(t)}$.
\end{theorem}

\subsubsection{Jump Sampling}
Standard reservoir sampling incurs an $O(t)$ cost by generating a random number for every incoming sample $x_t$. However, as $t \to \infty$, the selection probability $p_t = m/t$ vanishes, leading to significant computational redundancy. To mitigate this, we introduce \textit{jump sampling}. Specifically, let $t$ denote the index of the most recent update; instead of sequential testing, we determine the next update time $t + \Delta + 1$ by directly sampling the jump length $\Delta \in \{0, 1, \dots\}$. Under this scheme, the algorithm remains entirely idle between two consecutive updates, effectively bypassing all intermediate non-selected samples.
\begin{theorem}[Distribution of Jump Length]\label{theorem:d8ms}
The survival function of the jump length $\Delta$ at update time $t$ is given by:
\begin{equation}\label{eq:msu8}
S(\delta) = P(\Delta > \delta) = \left( \frac{t}{t+\delta} \right)^m.
\end{equation}
\end{theorem}
Intuitively, as $t$ increases, $S(\delta)$ decays more slowly, implying that the expected jump length $\Delta$ grows as the stream progresses. Leveraging this closed-form survival function, we can efficiently compute $\Delta$ for discrete indices via inverse transform sampling:
\begin{equation}\label{eq:nxy6}
\Delta = \lfloor t \cdot (U^{-1/k} - 1) \rfloor, \quad U \sim \text{Uniform}(0,1).
\end{equation}
Based on the above theoretical derivation, we propose the Jump Reservoir Sampling (JRS) algorithm. The algorithm leverages Inverse Transform Sampling to directly generate the skip gap for the next reservoir update, thereby avoiding item-by-item processing of samples likely to be discarded. 
\begin{lemma}[Complexity of Jump Sampling]\label{lamma:d9dk}
The expected time complexity of reservoir sampling with jump sampling is:
\begin{equation}\label{eq:9dmj}
O\left( m \left( 1 + \log \frac{t}{m} \right) \right).
\end{equation}
\end{lemma}
Note that since $t \gg m$, Equation (\ref{eq:9dmj}) can be approximated as $O(m\log t)$. The derivations of Equations (\ref{eq:msu8}), (\ref{eq:nxy6}), and (\ref{eq:9dmj}) are provided in Appendix \ref{appendix:dom2}, Appendix \ref{appendix:9mmd}, and Appendix \ref{appendix:dcc6}, respectively. The complete algorithmic procedure for Jump Reservoir Sampling (JRS) is presented in Appendix \ref{appendix:di88}.

\subsection{Quantile Space Interpolation}\label{sec:mmmd}
DAE encodes numerical features via Euclidean distances between input $x$ and quantiles $\{q_1, \dots, q_m\}$, potentially overlooking underlying data density (Appendix \ref{appendix:00kd}). To address this, we propose encoding $x$ in the quantile space rather than the value space through a two-step process.

\subsubsection{Sample-to-Quantile Mapping}
We select $M-1$ quantiles $\mathcal{Q} = \{q_1, \dots, q_{M-1}\}$ from the reservoir $\mathcal{R}$ as estimates of the population quantiles, partitioning the population distribution into $M$ equi-probable bins, where $M \ll m$. For any sample $x \in [q_j, q_{j+1})$, we invoke the local uniformity assumption to derive a refined quantile estimate $\hat{p} \in [0, 1]$ by combining the absolute position of the bin with the relative position of $x$ within it:
\begin{equation}
\hat{p} = {\frac{j}{M}} + {\frac{1}{M} \cdot \frac{x - q_j}{q_{j+1} - q_j}}\text{,}
\end{equation}
where $j \in \{0, \dots, M-1\}$ denotes the number of bins fully preceded by $x$. The first term represents the cumulative frequency up to the $j$-th bin while the second captures its precise relative offset. To handle edge cases and prevent index overflow, boundary values $q_0 = x_{\min}$ and $q_M = x_{\max}$ must be predefined, ensuring the sample $x$ is mapped consistently within the support of the estimated distribution.

\subsubsection{Quantile Space Interpolation}
We then map $\hat{p}$ to the coordinate space $v = \hat{p} \cdot M = j + w$, where $w = \frac{x - q_j}{q_{j+1} - q_j}$ is the interpolation weight for the $(j+1)$-th bin. To capture the cumulative density, we define an $M$-dimensional ``thermometer-style'' vector $\mathbf{v}_{\mathrm{raw}}$ by fully activating the first $j$ indices and assigning the fractional weight $w$ to the $(j+1)$-th entry:
\begin{equation}
\mathbf{v}_{\mathrm{raw}} = [\underbrace{1, \dots, 1}_{j}, \frac{x - q_j}{q_{j+1} - q_j}, \underbrace{0, \dots, 0}_{M-j-1}]^\top
\end{equation}
This encoding smoothly preserves the ordinal relationships and cumulative properties inherent in the quantile space.

\subsubsection{Efficiency Analysis}
Quantile estimation requires sorting the reservoir samples, but re-sorting is only needed when the reservoir is updated. By Lemma \ref{lamma:d9dk}, the number of updates is logarithmic, and when $t \gg m$, the reservoir remains nearly unchanged. Thus, caching can be used to avoid repeated sorting. In contrast, DAE relies on order-statistic–based estimation and requires sorting each batch. Therefore, our method is substantially more efficient than DAE.

\subsection{Field-Aware Meta-Embedding Aggregation}\label{sec:mmqu}
While $\mathbf{v}_{\mathrm{raw}}$ captures global marginal distributions, numerical semantics are often context-dependent. Instead of a naive, per-field distribution approach that triggers a combinatorial explosion, we propose an adaptive modulation mechanism that transforms $\mathbf{v}_{\mathrm{raw}}$ into a context-specific \emph{conditional distribution} based on field embeddings $\mathbf{e}_f$. Here, $\mathbf{e}_f$ typically represents categorical features that govern the distribution of numerical features.

\subsubsection{Distribution Modulation}
We explore two strategies for generating the modulated weights.

\textbf{Affine Transformation.}
We derive a modulation matrix $\mathbf{W}_{\mathrm{mod}} \in \mathbb{R}^{M \times M}$ via $\mathbf{W}_{\mathrm{mod}} = \text{Reshape}(\mathbf{W}_{\mathrm{tran}} \mathbf{e}_f)$, where $\mathbf{W}_{\mathrm{tran}} \in \mathbb{R}^{M^2 \times d}$ is a learnable projection matrix, and compute:
\begin{equation}
\mathbf{w}_{\mathrm{tran}} = \beta \cdot \text{Sigmoid}(\mathbf{W}_{\mathrm{mod}} \mathbf{v}_{\mathrm{raw}}) + (1-\beta) \cdot \mathbf{v}_{\mathrm{raw}},
\end{equation}
enabling the model to adaptively reweight quantile ranges based on $\mathbf{e}_f \in \mathbb{R}^d$.

\textbf{Gating Mechanism.}
A gating vector $\mathbf{w}_{\mathrm{mod}} \in \mathbb{R}^M$ is computed as $\mathbf{w}_{\mathrm{mod}} = \text{Sigmoid}(\mathbf{W}_{\mathrm{gate}} \mathbf{e}_f)$ and applied via a Hadamard product, where $\mathbf{W}_{\mathrm{gate}} \in \mathbb{R}^{M \times d}$ is a learnable weight matrix:
\begin{equation}
\mathbf{w}_{\mathrm{gate}} = \beta \cdot (\mathbf{v}_{\mathrm{raw}} \odot \mathbf{w}_{\mathrm{mod}}) + (1-\beta) \cdot \mathbf{v}_{\mathrm{raw}},
\end{equation}
allowing the model to dynamically mask or amplify specific quantile intervals based on $\mathbf{e}_f$.

Both strategies yield the modulated weight vector $\mathbf{w} \in \mathbb{R}^M$ through a weighted sum of $\mathbf{v}_{\mathrm{raw}}$ and its transformed counterpart. Here, $\beta \in [0, 1]$ serves as a hyperparameter to control the influence of field-specific context.

\subsubsection{Meta-Embedding Aggregation}
The final representation is derived as a weighted sum of learnable meta-embeddings $\mathbf{E} \in \mathbb{R}^{M \times d}$: 
\begin{equation}
\mathbf{e}_{\mathrm{final}} = \mathbf{w}^\top \mathbf{E} = \sum_{i=1}^{M} w_i \mathbf{E}_i\text{.} 
\end{equation}

\section{Experiments}

\begin{table*}[t]
\centering
\caption{Performance comparison of different numerical feature embedding schemes across multiple benchmark datasets and CTR backbones. DAES shows statistically significant improvements over the best baseline.}\label{tab:cncc}
\resizebox{\textwidth}{!}{
\begin{tabular}{cccccccccccccccc}
\toprule
\multirow{2}{*}{Dataset} & \multirow{2}{*}{Group} & \multirow{2}{*}{Method} & \multicolumn{2}{c}{FNN} & \multicolumn{2}{c}{Wide \& Deep} & \multicolumn{2}{c}{DeepFM} & \multicolumn{2}{c}{IPNN} & \multicolumn{2}{c}{DCN v2} & \multicolumn{2}{c}{xDeepFM} & \multirow{2}{*}{Avg Rank} \\
& & & AUC & LogLoss & AUC & LogLoss & AUC & LogLoss & AUC & LogLoss & AUC & LogLoss & AUC & LogLoss & \\
\midrule
\multirow{16}{*}{AutoML}
& \multirow{4}{*}{Static Binning} & EFD & 0.8258 & 0.0693 & 0.8284 & 0.0691 & 0.8194 & 0.0698 & 0.8273 & 0.0692 & 0.8255 & 0.0695 & 0.8266 & 0.0696 & 8.67 \\
& & LD & 0.8285 & 0.0692 & 0.8287 & 0.0692 & 0.8250 & 0.0692 & 0.8268 & 0.0693 & 0.8228 & 0.0697 & 0.8217 & 0.0701 & 9.00 \\
& & TD & 0.8257 & 0.0694 & 0.8261 & 0.0694 & 0.8251 & 0.0696 & 0.8264 & 0.0693 & 0.8268 & 0.0695 & 0.8253 & 0.0695 & 9.50 \\
& & MultiHot & 0.8283 & 0.0692 & 0.8279 & 0.0692 & 0.8268 & 0.0693 & 0.8288 & 0.0691 & 0.8195 & 0.0700 & 0.8267 & 0.0696 & 7.67 \\
\cmidrule{2-16}
& \multirow{7}{*}{Neural Embedding} & YouTube & 0.8217 & 0.0699 & 0.8226 & 0.0698 & 0.8204 & 0.0700 & 0.8235 & 0.0697 & 0.8221 & 0.0696 & 0.8228 & 0.0695 & 13.83 \\
& & DLRM & 0.8251 & 0.0695 & 0.8245 & 0.0696 & 0.8234 & 0.0698 & 0.8239 & 0.0697 & 0.8238 & 0.0693 & 0.8236 & 0.0694 & 12.33 \\
& & FE & 0.8268 & 0.0691 & 0.8262 & 0.0692 & 0.8271 & 0.0691 & 0.8255 & 0.0692 & 0.8213 & 0.0698 & 0.8084 & 0.0714 & 10.83 \\
& & AutoDis & 0.8263 & 0.0692 & 0.8292 & 0.0691 & 0.8247 & 0.0694 & 0.8253 & 0.0694 & 0.8238 & 0.0694 & 0.8282 & 0.0693 & 8.83 \\
& & DEER & 0.8279 & 0.0692 & 0.8275 & 0.0693 & 0.8271 & 0.0693 & 0.8268 & 0.0694 & 0.8234 & 0.0694 & 0.8261 & 0.0695 & 8.17 \\
& & Key-Value & 0.8262 & 0.0691 & 0.8260 & 0.0691 & 0.8261 & 0.0692 & 0.8261 & 0.0692 & 0.8179 & 0.0699 & 0.8226 & 0.0699 & 11.83 \\
& & NaryDis & 0.8287 & 0.0691 & 0.8298 & 0.0690 & 0.8281 & 0.0692 & 0.8295 & 0.0690 & 0.8251 & 0.0693 & 0.8284 & 0.0692 & 4.33 \\
\cmidrule{2-16}
& Interpolated Binning & PLE & 0.8302 & 0.0690 & 0.8317 & 0.0688 & 0.8308 & 0.0689 & 0.8317 & 0.0689 & 0.8253 & 0.0698 & 0.8281 & 0.0692 & 3.83 \\
\cmidrule{2-16}
& \multirow{3}{*}{\shortstack{Dynamic Quantile \\ Embedding}} & DAE & 0.8271 & 0.0693 & 0.8277 & 0.0692 & 0.8251 & 0.0696 & 0.8268 & 0.0694 & 0.8262 & 0.0695 & 0.8256 & 0.0695 & 8.17 \\
& & $\text{DAES}_{\text{gate}}$ & \textbf{0.8327} & \textbf{0.0688} & \textbf{0.8343} & \textbf{0.0686} & \textbf{0.8339} & \textbf{0.0686} & \textbf{0.8347} & \textbf{0.0685} & \textbf{0.8286} & \textbf{0.0692} & \textbf{0.8302} & \textbf{0.0691} & \textbf{1.33} \\
& & $\text{DAES}_{\text{tran}}$ & \textbf{0.8323} & \textbf{0.0688} & \textbf{0.8332} & \textbf{0.0687} & \textbf{0.8324} & \textbf{0.0687} & \textbf{0.8325} & \textbf{0.0687} & \textbf{0.8288} & \textbf{0.0692} & \textbf{0.8339} & \textbf{0.0687} & \textbf{1.67} \\
\midrule
\multirow{16}{*}{Criteo}
& \multirow{4}{*}{Static Binning} & EFD & 0.7858 & 0.4659 & 0.7858 & 0.4664 & 0.7858 & 0.4663 & 0.7859 & 0.4661 & 0.7857 & 0.4665 & 0.7823 & 0.4664 & 7.67 \\
& & LD & 0.7840 & 0.4670 & 0.7839 & 0.4670 & 0.7840 & 0.4669 & 0.7841 & 0.4670 & 0.7836 & 0.4671 & 0.7803 & 0.4715 & 11.50 \\
& & TD & 0.7835 & 0.4668 & 0.7834 & 0.4674 & 0.7832 & 0.4673 & 0.7831 & 0.4707 & 0.7839 & 0.4669 & 0.7837 & 0.4686 & 12.00 \\
& & MultiHot & 0.7861 & 0.4660 & 0.7864 & 0.4662 & 0.7861 & 0.4662 & 0.7860 & 0.4658 & 0.7858 & 0.4665 & 0.7849 & 0.4663 & 5.33 \\
\cmidrule{2-16}
& \multirow{7}{*}{Neural Embedding} & YouTube & 0.7835 & 0.4672 & 0.7821 & 0.4681 & 0.7824 & 0.4680 & 0.7829 & 0.4705 & 0.7832 & 0.4673 & 0.7839 & 0.4682 & 13.17 \\
& & DLRM & 0.7832 & 0.4672 & 0.7839 & 0.4672 & 0.7831 & 0.4674 & 0.7838 & 0.4695 & 0.7836 & 0.4671 & 0.7834 & 0.4689 & 12.67 \\
& & FE & 0.7848 & 0.4665 & 0.7846 & 0.4667 & 0.7848 & 0.4667 & 0.7846 & 0.4667 & 0.7845 & 0.4668 & 0.7849 & 0.4667 & 9.17 \\
& & AutoDis & 0.7864 & 0.4672 & 0.7865 & 0.4665 & 0.7838 & 0.4682 & 0.7843 & 0.4692 & 0.7868 & 0.4668 & 0.7859 & 0.4662 & 6.17 \\
& & DEER & 0.7852 & 0.4677 & 0.7859 & 0.4664 & 0.7858 & 0.4662 & 0.7851 & 0.4679 & 0.7856 & 0.4663 & 0.7854 & 0.4670 & 7.17 \\
& & Key-Value & 0.7626 & 0.4836 & 0.7629 & 0.4835 & 0.7627 & 0.4835 & 0.7628 & 0.4833 & 0.7618 & 0.4842 & 0.7610 & 0.4839 & 15.00 \\
& & NaryDis & 0.7860 & 0.4669 & 0.7865 & 0.4662 & 0.7858 & 0.4661 & 0.7857 & 0.4663 & 0.7862 & 0.4667 & 0.7855 & 0.4675 & 5.67 \\
\cmidrule{2-16}
& Interpolated Binning & PLE & 0.7877 & 0.4638 & 0.7879 & 0.4640 & 0.7877 & 0.4641 & 0.7876 & 0.4638 & 0.7872 & 0.4645 & 0.7879 & 0.4649 & 3.00 \\
\cmidrule{2-16}
& \multirow{3}{*}{\shortstack{Dynamic Quantile \\ Embedding}} & DAE & 0.7859 & 0.4671 & 0.7847 & 0.4670 & 0.7854 & 0.4665 & 0.7845 & 0.4686 & 0.7851 & 0.4668 & 0.7842 & 0.4680 & 8.50 \\
& & $\text{DAES}_{\text{gate}}$ & \textbf{0.7892} & \textbf{0.4625} & \textbf{0.7895} & \textbf{0.4625} & \textbf{0.7894} & \textbf{0.4624} & \textbf{0.7890} & \textbf{0.4633} & \textbf{0.7891} & \textbf{0.4626} & \textbf{0.7894} & \textbf{0.4622} & \textbf{1.50} \\
& & $\text{DAES}_{\text{tran}}$ & \textbf{0.7893} & \textbf{0.4625} & \textbf{0.7895} & \textbf{0.4622} & \textbf{0.7896} & \textbf{0.4622} & \textbf{0.7895} & \textbf{0.4623} & \textbf{0.7891} & \textbf{0.4625} & \textbf{0.7888} & \textbf{0.4631} & \textbf{1.50} \\
\midrule
\multirow{16}{*}{Industrial}
& \multirow{4}{*}{Static Binning} & EFD & 0.7556 & 0.2657 & 0.7559 & 0.2663 & 0.7559 & 0.2662 & 0.7557 & 0.2620 & 0.7558 & 0.2664 & 0.7522 & 0.2663 & 7.67 \\
& & LD & 0.7534 & 0.2667 & 0.7535 & 0.2672 & 0.7535 & 0.2672 & 0.7532 & 0.2708 & 0.7538 & 0.2668 & 0.7537 & 0.2683 & 12.50 \\
& & TD & 0.7542 & 0.2674 & 0.7538 & 0.2672 & 0.7541 & 0.2668 & 0.7543 & 0.2672 & 0.7538 & 0.2671 & 0.7502 & 0.2675 & 11.67 \\
& & MultiHot & 0.7564 & 0.2662 & 0.7565 & 0.2661 & 0.7562 & 0.2663 & 0.7562 & 0.2656 & 0.7556 & 0.2667 & 0.7547 & 0.2662 & 5.67 \\
\cmidrule{2-16}
& \multirow{7}{*}{Neural Embedding} & YouTube & 0.7534 & 0.2671 & 0.7523 & 0.2682 & 0.7523 & 0.2681 & 0.7527 & 0.2705 & 0.7533 & 0.2674 & 0.7538 & 0.2683 & 14.00 \\
& & DLRM & 0.7534 & 0.2673 & 0.7538 & 0.2671 & 0.7533 & 0.2675 & 0.7537 & 0.2696 & 0.7537 & 0.2673 & 0.7535 & 0.2686 & 13.33 \\
& & FE & 0.7557 & 0.2673 & 0.7546 & 0.2671 & 0.7553 & 0.2667 & 0.7546 & 0.2687 & 0.7552 & 0.2667 & 0.7541 & 0.2684 & 8.50 \\
& & AutoDis & 0.7553 & 0.2676 & 0.7558 & 0.2665 & 0.7557 & 0.2669 & 0.7553 & 0.2678 & 0.7558 & 0.2663 & 0.7555 & 0.2671 & 7.00 \\
& & DEER & 0.7565 & 0.2674 & 0.7566 & 0.2666 & 0.7537 & 0.2683 & 0.7545 & 0.2692 & 0.7569 & 0.2669 & 0.7558 & 0.2661 & 6.00 \\
& & Key-Value & 0.7535 & 0.2673 & 0.7538 & 0.2673 & 0.7534 & 0.2677 & 0.7538 & 0.2695 & 0.7538 & 0.2672 & 0.7534 & 0.2688 & 12.67 \\
& & NaryDis & 0.7562 & 0.2668 & 0.7564 & 0.2663 & 0.7557 & 0.2663 & 0.7558 & 0.2665 & 0.7562 & 0.2667 & 0.7555 & 0.2674 & 5.83 \\
\cmidrule{2-16}
& Interpolated Binning & PLE & 0.7574 & 0.2647 & 0.7577 & 0.2641 & 0.7576 & 0.2643 & 0.7573 & 0.2648 & 0.7571 & 0.2647 & 0.7579 & 0.2647 & 3.00 \\
\cmidrule{2-16}
& \multirow{3}{*}{\shortstack{Dynamic Quantile \\ Embedding}} & DAE & 0.7544 & 0.2664 & 0.7547 & 0.2666 & 0.7547 & 0.2666 & 0.7545 & 0.2668 & 0.7546 & 0.2669 & 0.7549 & 0.2666 & 9.17 \\
& & $\text{DAES}_{\text{gate}}$ & \textbf{0.7594} & \textbf{0.2626} & \textbf{0.7596} & \textbf{0.2624} & \textbf{0.7593} & \textbf{0.2623} & \textbf{0.7593} & \textbf{0.2632} & \textbf{0.7593} & \textbf{0.2628} & \textbf{0.7594} & \textbf{0.2622} & \textbf{1.33} \\
& & $\text{DAES}_{\text{tran}}$ & \textbf{0.7593} & \textbf{0.2626} & \textbf{0.7596} & \textbf{0.2623} & \textbf{0.7597} & \textbf{0.2623} & \textbf{0.7595} & \textbf{0.2621} & \textbf{0.7593} & \textbf{0.2624} & \textbf{0.7587} & \textbf{0.2632} & \textbf{1.67} \\
\bottomrule
\end{tabular}
}
\end{table*}

\subsection{Settings}

\subsubsection{Datasets and Metrics}
We evaluate DAES on two public benchmarks (Criteo, AutoML) and one large-scale industrial dataset, all chronologically ordered by exposure time. We provide a detailed description of the datasets and offer evidence indicating that their numerical features exhibit non-stationary distributions in Appendix \ref{appendix:993k}.

Performance is measured via AUC and LogLoss. We report the average results over five independent runs and perform a two-tailed unpaired t-test to assess statistical significance.

\subsubsection{Baselines and Backbones}
We compare DAES with four categories of numerical representation methods, all of which are introduced in §\ref{sec:mmeq}: \textit{Static Binning} (EFD~\cite{qu2018product}, LD~\cite{juan2016field}, TD~\cite{ke2019deepgbm}, Multihot~\cite{zhangmultihot}), \textit{Neural Embedding} (YouTube~\cite{covington2016deep}, DLRM~\cite{naumov2019deep}, FE~\cite{gorishniy2021revisiting}, AutoDis~\cite{guo2021embedding}, NaryDis~\cite{chen2022numerical}, Key-Value~\cite{guo2019multi}, DEER~\cite{cheng2022dynamic}), \textit{Interpolated Binning} (PLE~\cite{gorishniy2022embeddings}), and \textit{Dynamic Quantile Embedding} (DAE~\cite{shen2023dae}).

To evaluate its versatility and generalizability, all methods are implemented across six backbone architectures: FNN~\cite{covington2016deep}, Wide \& Deep~\cite{cheng2016wide}, DeepFM~\cite{guo2017deepfm}, IPNN~\cite{qu2018product}, DCN v2~\cite{wang2021dcn}, and xDeepFM~\cite{lian2018xdeepfm}.

\subsubsection{Implementation Details}
All models are trained using the Adam optimizer~\cite{kingma2014adam} with a batch size of 8,192. We tune the learning rate and L2 regularization in $\{10^{-6}, \cdots, 10^{-3}\}$. Model selection is based on LogLoss on the validation set, with early stopping set to 5 rounds. The default embedding size is 16, with each numerical field equipped with 10 meta-embeddings, and the MLP architecture is $(128, 32, 8)$. Dropout is set to 0.2, and BatchNorm is applied to stabilize training. For DCN v2 and xDeepFM, the explicit feature interaction modules (i.e., CrossNet and CIN) are configured with 3 layers. For DAES, the reservoir size is $10^5$, $\beta = 0.5$, and the three smallest-cardinality categorical features are used for distribution modulation\footnote{For a fair comparison, we fix $\beta = 0.5$ to align the hyperparameter count with baseline methods, while its sensitivity is analyzed in §\ref{sec:kkks}. Feature selection for distribution modulation is further discussed in §\ref{sec:8uks}.}. We refer to DAES using affine transformation as $\text{DAES}_{\text{tran}}$ and DAES using gating mechanism as $\text{DAES}_{\text{gate}}$.

For methods that require distribution information and support online estimation, including DAES and DAE, we perform a single pass through the data for online estimation. For methods that rely on offline statistical distribution information, including static binning and interpolated binning, we conduct offline statistics. Note that this inconsistency actually favors the baseline methods.

\subsection{Overall Performance}
Performance comparisons across three benchmark datasets and various deep CTR backbones are summarized in Table \ref{tab:cncc}. Our key observations are as follows:

\textbf{Discretization significantly outperforms non-discretization methods.}
Non-parametric schemes (e.g., YouTubeDNN’s No-embedding) fail to capture complex distributions, while shared DNN projections (e.g., DLRM) overlook field heterogeneity. Although FE introduces linear scaling, its shared nature restricts flexibility. In contrast, discretization expands model capacity via field-specific meta-embeddings, providing the necessary discriminability for complex feature interactions.

\textbf{Static Binning and Neural Embedding exhibit comparable performance.}
While neural embedding theoretically allows for joint optimization and continuity, they often suffer from representation collapse due to low discriminability. Conversely, although static binning is non-smooth and optimized asynchronously, heuristic binning introduces valuable prior distribution information that mitigates the impact of non-smoothness.

\textbf{Interpolated binning demonstrates excellent performance.} PLE achieves the best results among all baselines, including neural embedding methods that feature end-to-end optimization and smooth embeddings. Its advantage comes from encoding distribution information based on offline statistics, which incorporates distribution knowledge while ensuring smoothness through interpolation. This underscores the importance of leveraging distribution information. Notably, methods like PLE rely on offline statistics rather than true end-to-end pipelines, which limits their adaptability to dynamic environments.

\textbf{DAE exhibits inferior performance compared to DAES.}
Although DAE shares key attributes with DAES—namely distributional modeling, end-to-end optimization, and embedding smoothing—it suffers from systematic bias in quantile estimation within streaming scenarios. This bias, originating from the expectation of order statistics, arises when the i.i.d. assumption is violated. Furthermore, the interpolation method employed by DAE in the value space compromises the quality of the learned representations.

\textbf{DAES achieves robust and consistently superior performance across all backbones.}
Unlike baselines that vary across datasets, DAES consistently achieves superior performance across all backbones, demonstrating robust compatibility. We attribute these gains to the introduction of distributional information and field-aware distribution modulation. Note that even marginal improvements in AUC can translate into substantial commercial value~\cite{cheng2016wide}.

\subsection{Online A/B Testing}
We conducted 7-day online A/B tests on a large-scale platform serving hundreds of millions of daily active users to evaluate our proposed DAES against the production baseline, which utilizes manual discretization rules (e.g., EFD, LD). Notably, while AutoDis was initially considered, it was not transitioned to online A/B testing as its offline performance fell short of the existing production baseline.

The online results demonstrate that DAES significantly outperformed the production baseline, achieving a 2.307\% increase in Average Revenue Per User (ARPU)\footnote{ARPU is the primary metric used to evaluate performance within our advertising system.}. These improvements, which were found to be statistically significant, underscore the practical value of DAES in driving commercial growth. Beyond performance gains, DAES streamlines feature engineering by eliminating the need for offline statistics or handcrafted rules for both existing and future features. Currently, DAES has been fully deployed across our advertising platform.

\subsection{Ablation Analysis}
\subsubsection{Distribution Estimation}
We evaluate the distribution-preserving capability of reservoir sampling (RS) on the Criteo dataset, focusing on its 3rd and 5th numerical features (I3 and I5). Ground-truth distributions are derived from offline statistics on the full data, partitioned into 100 intervals using the 1\%, 2\%, …, 99\% quantiles. The estimated distributions produced by different methods are then compared with the ground-truth distribution, and the error is measured using KL divergence. As shown in Table \ref{tab:iima}, the proposed RS method achieves significantly lower KL divergence than DAE, which estimates via the expectation of order statistics (OS). RS demonstrates superior robustness across both clustered and spread distributions\footnote{See Appendix \ref{appendix:99ut} for detailed analysis of how feature distribution characteristics affect the performance of different methods.}. Furthermore, jump reservoir sampling (JRS) maintains comparable accuracy while reducing random number generation to only 3\%–4\% of that in the original scheme.

\subsubsection{Interpolation Space}
\begin{figure}[t]
  \centering
  \includegraphics[width=0.9\linewidth]{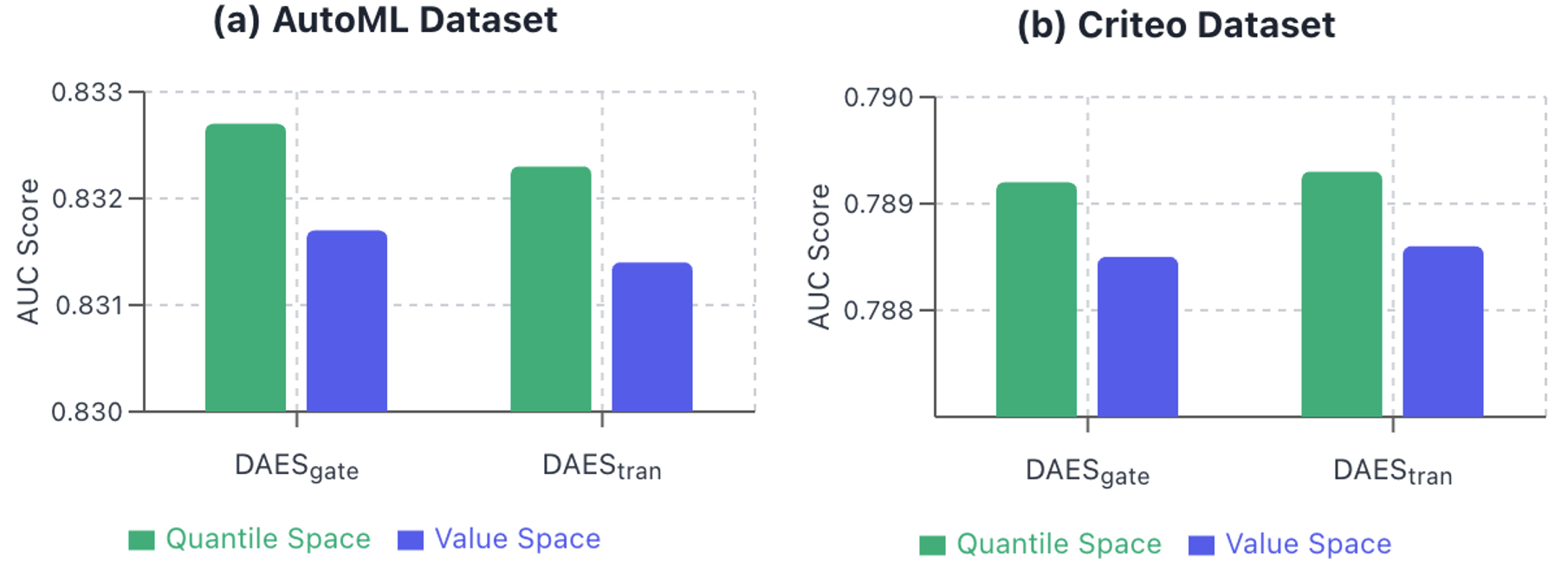}
  \caption{Performance comparison of interpolation conducted in value space versus quantile space.}\label{fig:mmd3}
  \Description{}
\end{figure}

Figure \ref{fig:mmd3} compares interpolation in the value versus quantile spaces. The results demonstrate that quantile-space interpolation yields superior performance.

\subsubsection{Distribution Modulation}\label{sec:kkks}
\begin{figure}[t]
  \centering
  \includegraphics[width=0.98\linewidth]{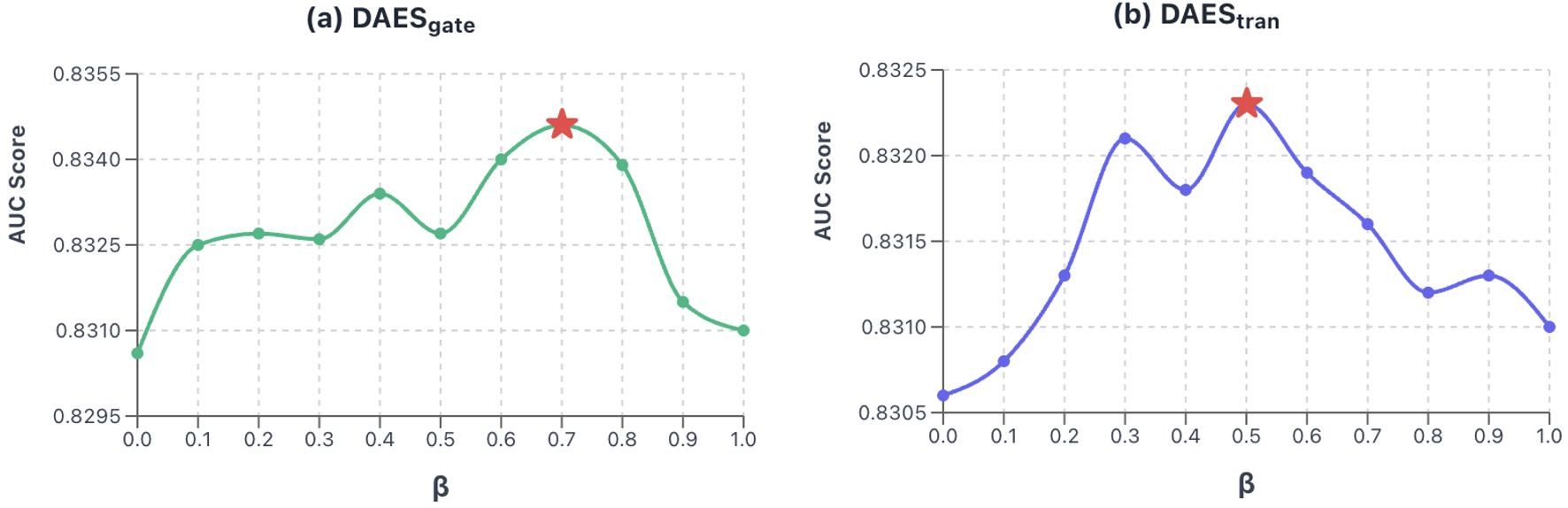}
  \caption{Impact of the modulation hyperparameter $\beta$ on model performance.}\label{fig:q9nv}
  \Description{}
\end{figure}

The impact of field-aware distribution modulation on model performance is regulated by adjusting the hyperparameter $\beta$. As illustrated in Figure \ref{fig:q9nv}, aggregating meta-embeddings with modulated weights significantly outperforms the unmodulated baseline, validating the effectiveness of incorporating contextual information at the embedding stage. In essence, DAES performs feature calibration at the embedding layer to augment its representational capacity. This approach reduces the reliance on downstream feature interaction modules to learn such calibrations, thereby alleviating the optimization complexity imposed on those subsequent layers.

\begin{table}[t]\footnotesize
\centering
\caption{Evaluation of distribution estimation methods for Criteo features I3 and I5 via KL divergence. Note: Sample counts are subject to minor variations due to NaNs.}\label{tab:iima}
\begin{tabular}{ccccc}
\toprule
\multirow{2}{*}{Method} & \multicolumn{2}{c}{I3} & \multicolumn{2}{c}{I5} \\
\cmidrule(lr){2-3} \cmidrule(lr){4-5}
& KL & \# Calls & KL & \# Calls \\
\midrule
OS  
& $5.97 \times 10^{0}$ 
& -- 
& $1.95 \times 10^{-1}$ 
& -- \\
RS
& $2.46 \times 10^{-4}$ 
& $2.93 \times 10^{7}$ 
& $4.54 \times 10^{-4}$ 
& $3.62 \times 10^{7}$ \\
JRS
& $1.60 \times 10^{-4}$ 
& $1.05 \times 10^{6}$ 
& $5.91 \times 10^{-4}$ 
& $1.09 \times 10^{6}$ \\
\bottomrule
\end{tabular}
\end{table}

\begin{figure}[t]
  \centering
  \includegraphics[width=0.98\linewidth]{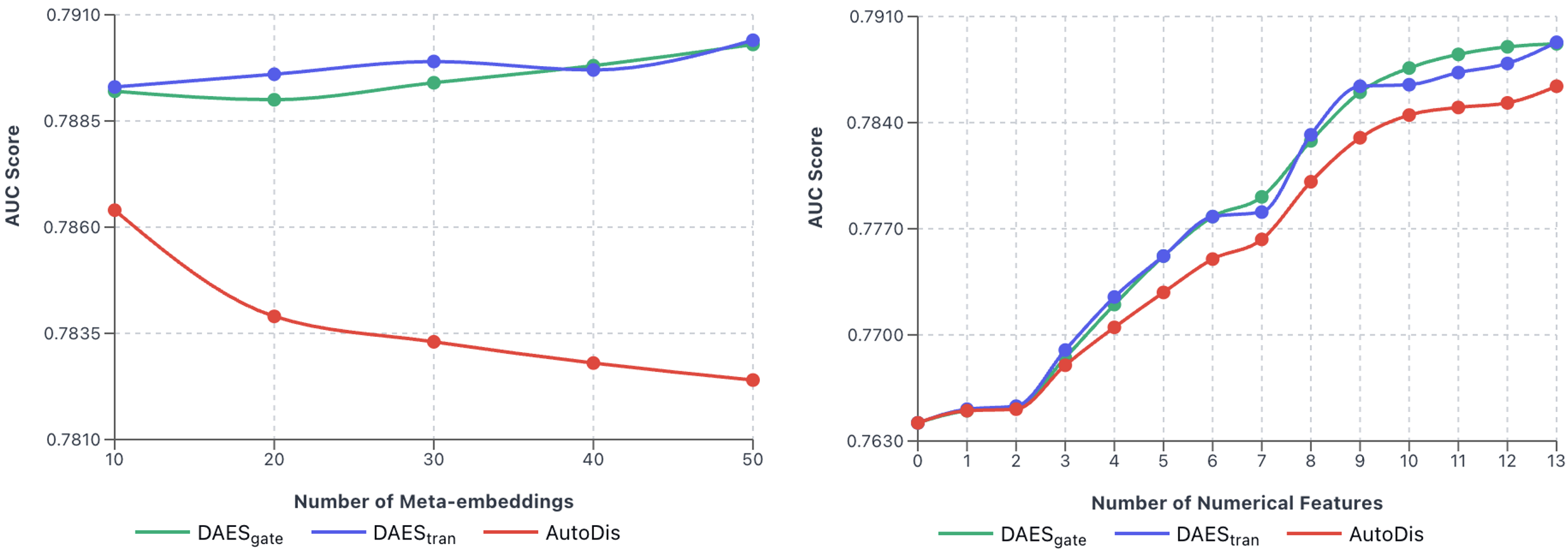}
  \caption{Performance analysis of DAES and AutoDis across varying meta-embedding counts and numbers of numerical features on the Criteo dataset.}\label{fig:932q}
  \Description{}
\end{figure}

\begin{figure}[t]
  \centering
  \includegraphics[width=0.98\linewidth]{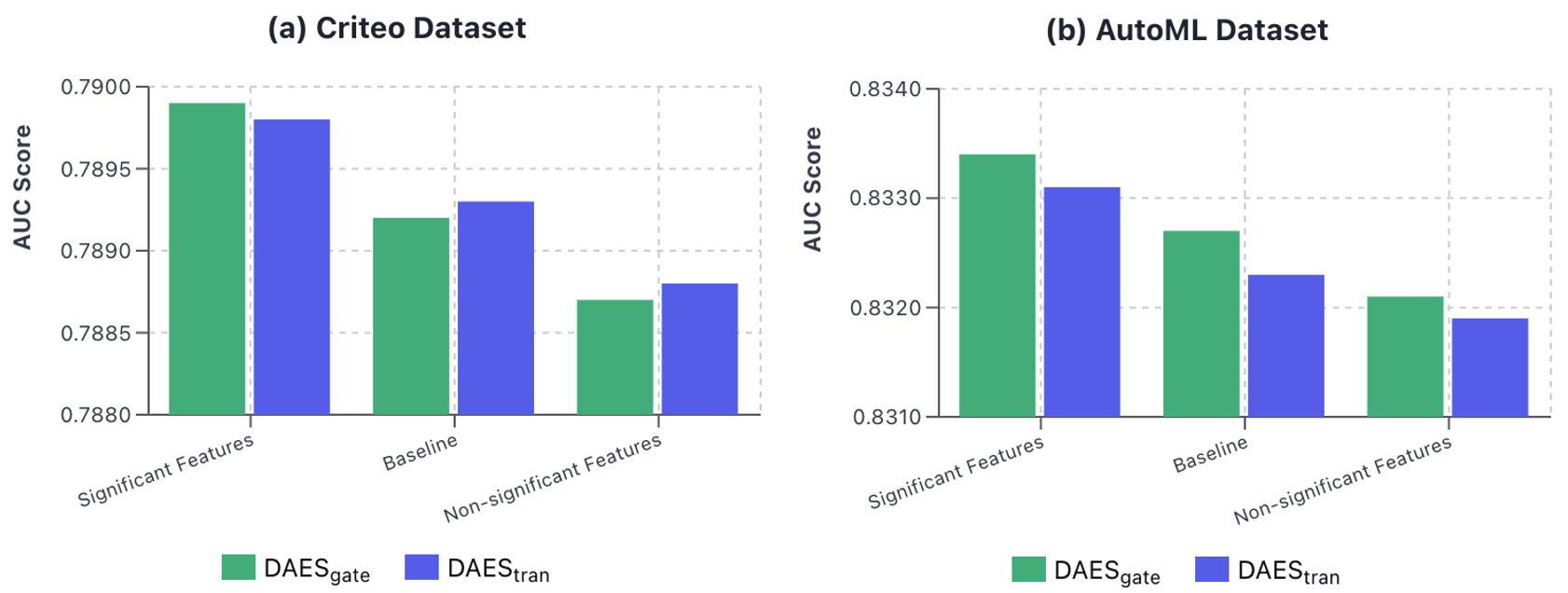}
  \caption{Performance comparison of distribution modulation using features selected by different criteria.}\label{fig:d98q}
  \Description{}
\end{figure}

\subsection{Analysis}

\subsubsection{Impact of Meta-embedding Count}
We investigated the impact of meta-embedding counts $M \in \{10, 20, \dots, 50\}$ on performance, using AutoDis as a representative end-to-end baseline. As shown in Figure \ref{fig:932q}, DAES consistently outperforms AutoDis. Notably, DAES is less sensitive to changes in the number of meta-embeddings and can even gain marginal improvements, whereas AutoDis exhibits the opposite trend. This confirms that while neural binning struggles to inherently capture data distributions, DAES leverages explicit distribution priors to guide and facilitate effective coarse-to-fine numerical representation learning.

\subsubsection{Scalability with Numerical Features}
Starting with 26 categorical fields from the Criteo dataset, we incrementally introduced 13 numerical fields. Figure \ref{fig:932q} shows that DAES consistently achieves performance gains with each newly added feature, outperforming AutoDis in all cases. These results demonstrate the robustness and stable improvement of DAES as more numerical features are incorporated.

\subsubsection{Feature Selection for Distribution Modulation}\label{sec:8uks}
Previously, we simply selected the three categorical features with the smallest cardinality as the baseline for distribution modulation. In this section, we first employ the Kruskal-Wallis test to assess whether a categorical feature significantly affects the conditional distribution of numerical features. For features passing the significance test, we further use the inter-group Wasserstein distance as an effect size to quantify the distributional difference strength, thereby identifying three significant and three non-significant features. Figure \ref{fig:d98q} demonstrates that employing more significant features for distribution modulation achieves superior performance.

\section{Conclusions}
This paper addresses numerical feature embedding in CTR models within streaming scenarios. By analyzing the limitations of current methods in distribution-awareness and context-dependency, we propose DAES, an end-to-end framework that dynamically models and modulates feature distributions. Empirical results confirm DAES's consistent superiority over state-of-the-art baselines. Methodologically, we demonstrate that incorporating distribution as a first-class modeling element provides a principled, scalable approach for online learning, with potential applications across various online systems facing non-stationary data distributions.

\begin{acks}
\end{acks}

\balance

\bibliographystyle{ACM-Reference-Format}
\bibliography{sample-base}


\appendix

\section{Dataset Introduction}\label{appendix:993k}

\begin{table}[t]\footnotesize
\centering
\caption{Dataset Statistics.}\label{tab:jmaa}
\begin{tabular}{cccc}
\toprule
Dataset & \# Cat Features & \# Num Features & \# Instances \\
\midrule
AutoML     & 25 & 6  & $8.7 \times 10^6$  \\
Criteo     & 26 & 13 & $45.8 \times 10^6$ \\
Industrial & 20 & 10 & $87.3 \times 10^6$ \\
\bottomrule
\end{tabular}
\end{table}

\begin{figure*}[t]
  \centering
  \includegraphics[width=0.98\linewidth]{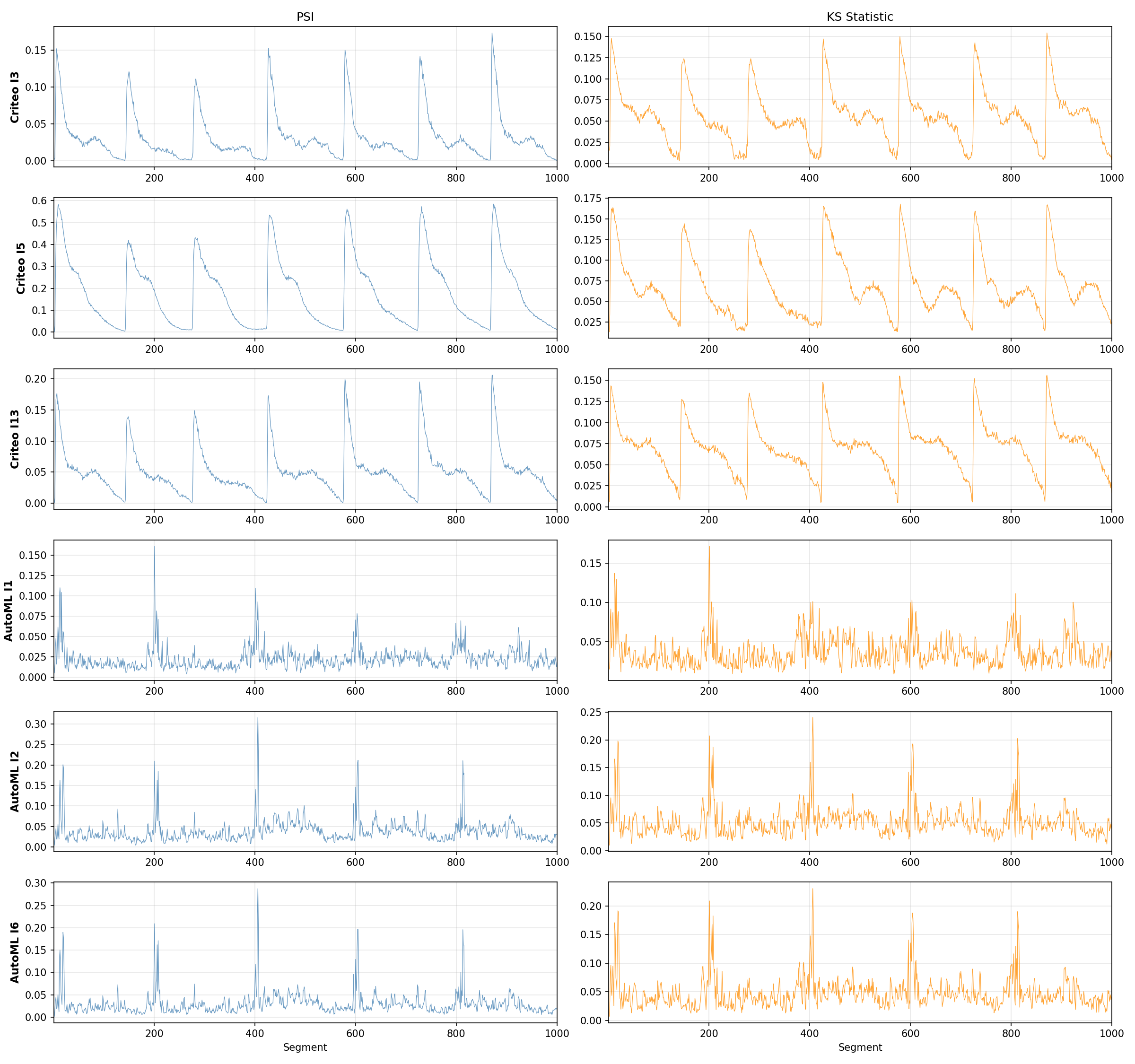}
  \caption{PSI and KS statistics over time for selected numerical features, demonstrating periodic distribution shifts in Criteo and AutoML datasets.}\label{fig:d9zs}
  \Description{}
\end{figure*}

We evaluate our method on three representative datasets with different characteristics, as described below. Dataset statistics are summarized in Table \ref{tab:jmaa}.

\begin{itemize}
    \item \textbf{Criteo\footnote{\url{https://www.kaggle.com/c/criteo-display-ad-challenge}}:} Released as part of the Criteo Display Advertising Challenge 2013, this dataset is widely used for CTR model evaluation and contains 13 numerical feature fields.
    \item \textbf{AutoML\footnote{\url{https://www.4paradigm.com/competition/nips2018}}:} Originating from the NeurIPS 2018 “AutoML for Lifelong Machine Learning” Challenge, it contains 6 numerical feature fields.
    \item \textbf{Industrial:} Sampled and collected from a major online advertising platform, containing 10 numerical feature fields.
\end{itemize}
For the public datasets Criteo and AutoML, we partition the temporally ordered data into 1,000 segments and calculate their Population Stability Index (PSI) and Kolmogorov-Smirnov statistic (KS) against the first segment. PSI and KS can be used to measure distributional similarity, where larger PSI and KS values indicate greater distributional divergence. As illustrated in Figure \ref{fig:d9zs}, where we present only a subset of numerical features (I3, I5, I13 for Criteo and I1, I2, I6 for AutoML), the distributional similarity exhibits fluctuations and demonstrates periodic patterns, suggesting that data distributions in streaming scenarios violate the independent and identically distributed (i.i.d.) assumption.

\section{Theoretical Limitation of Order Statistics in Non-IID Streams}\label{appendix:93km}
DAE~\cite{shen2023dae} leverages the expectation of order statistics to estimate population quantiles. However, this methodology relies heavily on the assumption of Independent and Identically Distributed (IID) data, which is frequently violated in real-world streaming scenarios characterized by non-stationary distributions. In such contexts, local estimates derived from individual batches become biased.

To illustrate why traditional order statistics fail under Non-IID conditions, consider a population distribution $X$ following a uniform distribution $U[a, b]$. In a streaming framework, assume the data is partitioned into $t$ temporal intervals (batches), where the $i$-th interval follows a local uniform distribution $U[a_i, b_i]$ for $i=1, 2, \dots, t$. We define the boundaries such that the data shifts linearly over time:
\begin{equation}
a_i = a + \frac{i-1}{t}(b-a), \quad b_i = a + \frac{i}{t}(b-a)
\end{equation}
In this non-stationary stream, the local $\alpha$-quantile for the $i$-th batch, denoted as $q_{i, \alpha}$, is given by:
\begin{equation}
q_{i, \alpha} = a_i + \alpha(b_i - a_i) = a + \frac{i-1+\alpha}{t}(b-a)
\end{equation}
The expectation of these local $\alpha$-quantiles across the entire stream is calculated as:
\begin{equation}
\begin{aligned}
E[q_{\alpha}] &= \frac{1}{t} \sum_{i=1}^{t} \left[ a + \frac{i-1+\alpha}{t}(b-a) \right] \\
&= a + \frac{b-a}{t^2} \left[ \sum_{i=1}^{t}(i-1) + \sum_{i=1}^{t}\alpha \right] \\
&= a + (b-a) \cdot \frac{t-1+2\alpha}{2t}
\end{aligned}
\end{equation}
In contrast, the true global $\alpha$-quantile of the population distribution $U[a, b]$ is:
\begin{equation}
Q_{\alpha} = a + \alpha(b-a)
\end{equation}
Comparing the two results, we observe that:
\begin{equation}
E[q_{\alpha}] - Q_{\alpha} = \frac{(t-1)(1-2\alpha)}{2t}(b-a)
\end{equation}
It is evident that $E[q_{\alpha}] \neq Q_{\alpha}$ unless $t=1$ (the IID case) or $\alpha=0.5$ (the median in a symmetric shift). Consequently, the expectation of local order statistics is an asymptotically biased estimator of the global quantile in the presence of concept drift. This discrepancy arises because local batches fail to capture the global support of the distribution, leading to a systematic failure of DAE to converge to the true distributional characteristics of non-stationary data streams.

\section{Limitations of DAE in Value-Space Encoding}\label{appendix:00kd}
DAE~\cite{shen2023dae} operates by encoding numerical values within the value space. Specifically, given $m$ quantiles $q_1, \dots, q_m$, DAE generates an $m$-dimensional feature vector by calculating the numerical distances $|x - q_i|$ between the input $x$ and each quantile. In regions characterized by long-tail distributions or abrupt density variations, this approach is susceptible to topological errors. Such errors occur when a sample $x$ statistically belongs to the tail of Bucket A, yet its numerical representation exhibits a closer proximity to the head of Bucket B, thereby undermining the ordinal consistency of the learned features.

The root of this issue lies in the decoupling of physical span from probabilistic distance: in sparse regions, excessive numerical distances can overshadow local signals, and linear distances fail to capture the non-linear dynamics of the Cumulative Distribution Function (CDF). For instance, consider a long-tail distribution with quantiles $q_1=0$, $q_2=100$, and $q_3=105$. For an input $x=95$, its statistical membership lies within the interval $[q_1, q_2]$ (Bucket A). However, the resulting numerical distances $|x-q_2|=5$ and $|x-q_3|=10$ are both significantly smaller than the distance to its own bucket's origin, $|x-q_1|=95$. Consequently, the feature vector erroneously suggests a high affinity between $x$ and the subsequent Bucket B ($[q_2, q_3]$), causing the model to misidentify the sample's distributional context and fail to accurately perceive its relative position within the long-tail interval.

\section{Reservoir Sampling}\label{appendix:d9mp}
Let $m$ denote the fixed capacity of the reservoir $\mathcal{R}$. The sampling process is divided into two distinct phases.

\textbf{1) Initialization Phase ($t \le m$).} For the initial $m$ samples arriving in the stream, each sample is directly admitted into the reservoir without competition. At any time step $t \le m$, the state of the reservoir is defined as:
\begin{equation}
\mathcal{R}_t = \{x_1, x_2, \dots, x_t\}
\end{equation}

\textbf{2) Update Phase ($t > m$).} For each subsequent sample $x_t$ arriving at time $t$, the algorithm performs a probabilistic update to maintain a uniform random sample of all items seen so far:
\begin{itemize}
\item \textit{Acceptance.} The new sample $x_t$ is accepted into the reservoir with a probability of $P(\text{accept}) = \frac{m}{t}$.
\item \textit{Replacement.} If $x_t$ is accepted, one existing element in $\mathcal{R}$ is selected uniformly at random and replaced by $x_t$.
\item \textit{Discarding.} If not accepted, $x_t$ is discarded, and the reservoir remains unchanged.
\end{itemize}

\textbf{Computational Complexity:} The total computational complexity grows linearly with the number of samples $t$, denoted as $O(t)$, as each individual update requires exactly one random number generation.

\section{Proof of Equation (\ref{eq:d9mg})}\label{appendix:kkvz}
We prove by mathematical induction that at any time step $t$, the probability that any item $x_i$ (for $i \le t$) is contained in the reservoir $\mathcal{R}_t$ is exactly $m/t$.

\textbf{Base Case ($t = m$):}
At time step $t = m$, the reservoir $\mathcal{R}_m$ contains exactly the first $m$ samples, i.e., $\mathcal{R}_m = \{x_1, \dots, x_m\}$.
Therefore, for any $i \le m$,
\begin{equation}
P(x_i \in \mathcal{R}_m) = 1 = \frac{m}{m}.
\end{equation}
Hence, the base case holds.

\textbf{Inductive Hypothesis:}
Assume that the claim holds at time step $t-1$, namely, for any $i \le t-1$,
\begin{equation}
P(x_i \in \mathcal{R}_{t-1}) = \frac{m}{t-1}.
\end{equation}

\textbf{Inductive Step:}
At time step $t$, an existing sample $x_i$ (with $i < t$) remains in the reservoir if and only if:
\begin{enumerate}
\item it is present in the reservoir at the previous time step ($x_i \in \mathcal{R}_{t-1}$), and
\item it is not replaced by the newly arrived sample $x_t$.
\end{enumerate}
According to the reservoir sampling update rule, the new sample $x_t$ is accepted into the reservoir with probability $m/t$. Conditional on being accepted, it replaces a uniformly chosen existing element, so the probability that it replaces a specific item $x_i$ is $1/m$. Consequently, the probability that $x_i$ is replaced at time $t$ is
\begin{equation}
\begin{aligned}
P(x_i \text{ is replaced})
&= P(x_t \text{ is accepted}) \cdot P(x_i \text{ is selected} \mid x_t \text{ is accepted}) \\
&= \frac{m}{t} \cdot \frac{1}{m} = \frac{1}{t}.
\end{aligned}
\end{equation}
Therefore, the probability that $x_i$ remains in the reservoir at time $t$ is
\begin{equation}
\begin{aligned}
P(x_i \in \mathcal{R}_t)
&= P(x_i \in \mathcal{R}_{t-1}) \cdot \left(1 - P(x_i \text{ is replaced})\right) \\
&= \frac{m}{t-1} \cdot \left(1 - \frac{1}{t}\right) \\
&= \frac{m}{t-1} \cdot \frac{t-1}{t} = \frac{m}{t}.
\end{aligned}
\end{equation}
For the newly arrived sample $x_t$, the probability of being in the reservoir is exactly its acceptance probability, $m/t$. Hence, the claim holds for all $i \le t$, completing the inductive proof.

\section{Derivation of the Survival Function} \label{appendix:dom2}

To derive the distribution of the jump length $\Delta$, we examine the process from both discrete and continuous perspectives.

\paragraph{Discrete Perspective (Bernoulli Trials)}
In the discrete setting, for any sample $x_{t+i}$ ($i > 0$), the probability that it is rejected (i.e., not selected to enter the reservoir) is $1 - \frac{m}{t+i}$. The event $\Delta > \delta$ occurs if and only if all $\delta$ consecutive samples from index $t+1$ to $t+\delta$ are rejected. Assuming independence between trials, the probability is given by the product:
\begin{equation}
P(\Delta > \delta) = \prod_{i=1}^{\delta} \left( 1 - \frac{m}{t+i} \right).
\end{equation}

\paragraph{Continuous Approximation (Non-homogeneous Poisson Process)}
As $t$ becomes large, calculating the product becomes computationally expensive and analytically intractable. We approximate the discrete selection process using a continuous-time non-homogeneous Poisson process with an intensity function $\lambda(u)$ representing the instantaneous selection rate:
\begin{equation}
\lambda(u) = \frac{m}{u}, \quad u > t.
\end{equation}
According to the properties of a non-homogeneous Poisson process, the number of events (updates) occurring in the interval $(t, t+\delta]$, denoted by $N_{(t, t+\delta]}$, follows a Poisson distribution with parameter $\Lambda$:
\begin{equation}
\Lambda = \int_{t}^{t+\delta} \lambda(u) \, du = \int_{t}^{t+\delta} \frac{m}{u} \, du = m \ln\left( \frac{t+\delta}{t} \right).
\end{equation}
The condition $\Delta > \delta$ is equivalent to the event that zero updates occur in the interval $(t, t+\delta]$. Using the Poisson probability mass function $P(X=n) = \frac{\Lambda^n e^{-\Lambda}}{n!}$ for $n=0$:
\begin{equation}
P(\Delta > \delta) = P(N_{(t, t+\delta]} = 0) = e^{-\Lambda}.
\end{equation}
Substituting the expression for $\Lambda$:
\begin{equation}
P(\Delta > \delta) = e^{-m \ln\left( \frac{t+\delta}{t} \right)} 
                   = \left( \frac{t+\delta}{t} \right)^{-m} 
                   = \left( \frac{t}{t+\delta} \right)^m.
\end{equation}
This concludes the derivation of the survival function $S(\delta)$. 

\section{Derivation of Jump Length}\label{appendix:9mmd}

Based on the principle of inverse transform sampling, let $U$ be a random variable uniformly distributed on $(0, 1)$. Since the range of the survival function $S(\delta) = P(\Delta > \delta)$ is also $(0, 1)$, we set:
\begin{equation}
U = S(\delta) = \left( \frac{t}{t+\delta} \right)^k.
\end{equation}
To solve for $\delta$, we take the $-1/k$ power of both sides:
\begin{equation}
U^{-1/k} = \frac{t+\delta}{t} = 1 + \frac{\delta}{t}.
\end{equation}
Rearranging the terms to isolate $\delta$ yields:
\begin{equation}
\frac{\delta}{t} = U^{-1/k} - 1 \implies \delta = t \cdot (U^{-1/k} - 1).
\end{equation}
Since the stream indices must be discrete integers, we apply the floor function to obtain the final formula for the jump length:
\begin{equation}
\Delta = \lfloor t \cdot (U^{-1/k} - 1) \rfloor.
\end{equation}

\section{Proof of Lemma \ref{lamma:d9dk}}\label{appendix:dcc6}
The total time complexity of the algorithm is proportional to the total number of updates performed on the reservoir. For a data stream of length $t$ and a reservoir of size $m$, the probability that the $i$-th element (where $i > m$) is selected to enter the reservoir is $m/i$. Let $N$ denote the total number of updates. The expected number of updates $E[N]$ is given by:
\begin{equation}
E[N] = m + \sum_{i=m+1}^{t} \frac{m}{i} = m \left( 1 + \sum_{i=m+1}^{t} \frac{1}{i} \right).
\end{equation}
The summation above can be elegantly expressed as the difference between two harmonic numbers, $H_t$ and $H_m$:
\begin{equation}
\sum_{i=m+1}^{t} \frac{1}{i} = \sum_{i=1}^{t} \frac{1}{i} - \sum_{i=1}^{m} \frac{1}{i} = H_t - H_m.
\end{equation}
By applying the asymptotic expansion for harmonic numbers, $H_n = \sum_{i=1}^n \frac{1}{i} \approx \ln n + \gamma$ (where $\gamma$ is the Euler-Mascheroni constant), we can approximate the expected updates as:
\begin{equation}
E[N] \approx m \left( 1 + (\ln t + \gamma) - (\ln m + \gamma) \right) = m \left( 1 + \ln \frac{t}{m} \right).
\end{equation}
Consequently, the expected update complexity is $O\left( m \left( 1 + \log \frac{t}{m} \right) \right)$. This bound demonstrates that the complexity grows logarithmically with the stream length $t$, offering a significant improvement over the $O(t)$ complexity of standard reservoir sampling.

\section{Algorithm Implementation}\label{appendix:di88}

\begin{algorithm}[ht]
\caption{Jump Reservoir Sampling via Inverse Transform}
\label{alg:jump_reservoir}
\begin{algorithmic}[1]
\Require Data stream $\mathcal{S}$ (potentially infinite), reservoir size $m$
\Ensure A uniformly random sample set $\mathcal{R}$

\State Initialize reservoir $\mathcal{R} \leftarrow \{x_1, x_2, \ldots, x_m\}$
\State $t \leftarrow m$ \Comment{Current position in the stream}
\State Sample $U \sim \text{Uniform}(0, 1)$
\State $\Delta \leftarrow \lfloor t \cdot (U^{-1/m} - 1) \rfloor$ \Comment{Initial jump gap}
\State $t_{\text{next}} \leftarrow t + \Delta + 1$ \Comment{Next update position}

\While{stream $\mathcal{S}$ continues}
    \State $t \leftarrow t + 1$ \Comment{Advance to next element}
    \State Read element $x_t$ from stream $\mathcal{S}$
    
    \If{$t < t_{\text{next}}$}
        \State \textbf{continue} \Comment{Skip this element}
    \EndIf
    
    \State \textit{// Update occurs at position $t = t_{\text{next}}$}
    \State Sample $j \sim \text{Uniform}\{1, 2, \ldots, m\}$
    \State $\mathcal{R}[j] \leftarrow x_t$
    
    \State \textit{// Generate next jump gap}
    \State Sample $U \sim \text{Uniform}(0, 1)$
    \State $\Delta \leftarrow \lfloor t \cdot (U^{-1/m} - 1) \rfloor$
    \State $t_{\text{next}} \leftarrow t + \Delta + 1$
\EndWhile

\State \Return $\mathcal{R}$
\end{algorithmic}
\end{algorithm}

The algorithm operates in two phases. In the initialization phase (lines 1--5), we fill the reservoir with the first $m$ elements, initialize the position counter $t \leftarrow m$, and compute the first jump gap $\Delta$ to determine the next update position $t_{\text{next}} = m + \Delta + 1$. The main loop (lines 6--18) processes the stream incrementally:

\textbf{Stream traversal (lines 7--10).} At each iteration, the position counter $t$ is incremented by 1, and the current element $x_t$ is read from the stream. If $t < t_{\text{next}}$, the element is skipped and the loop continues to the next iteration.

\textbf{Reservoir update (lines 12--14).} When $t = t_{\text{next}}$, an update occurs. A uniformly random index $j \sim \text{Uniform}\{1, 2, \ldots, m\}$ is sampled, and the reservoir is updated by setting $\mathcal{R}[j] \leftarrow x_t$.

\textbf{Jump gap generation (lines 15--18).} Immediately after each update, a new jump gap is computed. We sample $U \sim \text{Uniform}(0, 1)$ and apply the inverse transform $\Delta \leftarrow \lfloor t \cdot (U^{-1/m} - 1) \rfloor$ based on the survival function $S(\delta) = \left(\frac{t}{t+\delta}\right)^m$ from Theorem \ref{theorem:d8ms}. The next update position is then set to $t_{\text{next}} \leftarrow t + \Delta + 1$.

This formulation clearly separates the update logic from gap computation, with $\Delta$ being recalculated only when an update occurs at position $t_{\text{next}}$.

\section{Distribution Characteristics and Method Performance}\label{appendix:99ut}
Due to their inherently different distributions, I3 and I5 yield different relative results between RS and OS methods. This is particularly evident on I3, where values are highly clustered on small integers—a characteristic prevalent in CTR tasks. In such cases, the resulting overlap of quantiles causes OS methods to suffer from systematic bucket misassignment due to minor estimation errors at these dense boundaries. In contrast, I5 features a more spread-out distribution with distinct quantile intervals, allowing OS to perform better, though it still lags behind RS. These observations demonstrate that RS maintains robust performance regardless of whether the feature distribution is clustered or spread out.

\section{More Discussions}
This section provides a deeper examination of DAES's design choices, theoretical underpinnings, and practical considerations for industrial deployment.

\textbf{Trade-off Between Global Distribution Estimation and Temporal Adaptivity.} Our reservoir sampling approach maintains global historical samples to estimate long-term distributional characteristics. In practical CTR prediction scenarios, numerical feature distributions exhibit periodic fluctuations at daily granularity (see Appendix \ref{appendix:993k}) with overall stability. Over-pursuing short-term distributions (e.g., hourly windows) would cause frequent and substantial quantile shifts, introducing system instability and degrading user experience. However, we acknowledge that global sampling may exhibit latency in scenarios with significant concept drift~\cite{gama2014survey}. Two practical optimization strategies warrant exploration: (1) designing expiration mechanisms that periodically discard distant samples and reset indices to control sampling probabilities, balancing stability and adaptivity; (2) leveraging historical data to populate the reservoir for cold-start during initial deployment, avoiding distributional estimation bias in the early stage.

\textbf{Reservoir Sampling vs. Specialized Streaming Quantile Algorithms.} While specialized algorithms such as KLL sketch~\cite{karnin2016optimal}, Greenwald-Khanna~\cite{greenwald2001space}, and t-digest~\cite{dunning2019computing} may offer superior memory efficiency, we choose reservoir sampling based on the following considerations: (1) End-to-end differentiability: Traditional quantile algorithms are designed for query scenarios, and their complex data structures (e.g., compression trees, summary structures) are difficult to integrate seamlessly with neural network gradient backpropagation; (2) Global distribution estimation vs. point queries: DAES requires the complete empirical distribution for interpolation and modulation, rather than merely querying specific quantiles; (3) Implementation simplicity: Reservoir sampling is straightforward to implement in deep learning frameworks with low maintenance costs. Future work may explore adapting sketch algorithms into differentiable frameworks.

\textbf{Theoretical Contribution of Jump Sampling.} Although the idea of jump-based reservoir sampling originates from Vitter's seminal work~\cite{vitter1985random}, our derivation adopts a different theoretical framework. Vitter's derivation is based on the geometric distribution of discrete Bernoulli trials, whereas we establish a continuous-time approximation through a non-homogeneous Poisson process, providing a more intuitive probabilistic interpretation and naturally deriving the inverse transform sampling formula.

\textbf{Bias-Variance Trade-off in Reservoir Size.} The reservoir size $m$ presents a fundamental trade-off between estimation accuracy and computational cost. Let $\hat{q}_\alpha^{(m)}$ denote the $\alpha$-quantile estimated from a reservoir of size $m$, and $q_\alpha$ denote the true population quantile. The estimation error decomposes into:
\begin{equation}
\mathbb{E}[(\hat{q}_\alpha^{(m)} - q_\alpha)^2] = \text{Bias}^2(\hat{q}_\alpha^{(m)}) + \text{Var}(\hat{q}_\alpha^{(m)})
\end{equation}
In streaming scenarios, bias arises primarily from two sources: (1) finite-sample approximation, the reservoir fails to cover the full support of the true distribution; (2) distributional non-stationarity, concept drift between historical samples and the current distribution. Variance is governed by sampling randomness. According to asymptotic theory for quantile estimation, under mild regularity conditions (e.g., $f(q_\alpha) > 0$ where $f$ is the density function), the variance decays at an $O(1/m)$ rate:
\begin{equation}
\text{Var}(\hat{q}_\alpha^{(m)}) \approx \frac{\alpha(1-\alpha)}{m \cdot f^2(q_\alpha)}
\end{equation}
This indicates that larger $m$ reduces estimation variance, at the expense of $O(m \log m)$ sorting overhead and linear memory footprint. Our theoretical analysis (Theorem \ref{th:93mv}) establishes consistency as $m \to \infty$, but the bias term under finite samples is difficult to characterize explicitly, as it depends heavily on the geometric properties of the true distribution (tail heaviness, multimodality, local smoothness) and temporal evolution patterns.
Empirically, $m = 10^5$ demonstrates robustness across three datasets, though selecting optimal $m$ for specific distributional characteristics may require cross-validation. A promising direction is adaptively adjusting $m$ based on online metrics of quantile stability (e.g., rate of change across successive updates).

\textbf{System Overhead and Industrial Deployment Trade-offs.} A notable limitation of DAES is that each numerical feature requires maintaining an independent reservoir (approximately $m \times 4$ bytes = 400KB per feature for $m=10^5$), resulting in non-negligible memory overhead when the number of features is large. Moreover, despite jump sampling drastically reducing random number generation (to 3-4\% of the standard approach, Table \ref{tab:iima}), sorting operations during reservoir updates still introduce additional latency. In our industrial deployment, QPS decreased by 1.8\% compared to the production baseline, but the overall ROI remained positive due to a 2.307\% improvement in ARPU. For scenarios with extreme throughput demands, one may consider reducing reservoir size or employing approximate sorting methods. Future work may explore cross-feature reservoir sharing or hierarchical sampling structures to further optimize memory efficiency.
\end{document}